\documentclass[prl,reprint,amssymb,amsmath,superscriptaddress,showpacs,twocolumn]{revtex4-1}

\usepackage[pdftex]{graphicx}
\usepackage[usenames]{color}
\usepackage{sidecap}

\begin{document}
\bibliographystyle{apsrev}


\title[P. Perna \textit{et al.}] {Interfacial exchange coupling induced chiral symmetry-breaking of Spin-Orbit effects}

\author{P. Perna}
\thanks{Corresponding author: paolo.perna@imdea.org}
\affiliation{IMDEA-Nanoscience, c/ Faraday, 9 Campus de
Cantoblanco, Madrid, 28049 Madrid, Spain}

\author{F. Ajejas}
\affiliation{IMDEA-Nanoscience, c/ Faraday, 9 Campus de Cantoblanco, Madrid, 28049 Madrid, Spain}
\affiliation{Departamento de Fisica de la Materia Condensada and Instituto "Nicol\'as Cabrera", Universidad Aut\'{o}noma de Madrid, 28049 Madrid, Spain}

\author{D. Maccariello}
\affiliation{IMDEA-Nanoscience, c/ Faraday, 9 Campus de Cantoblanco, Madrid, 28049 Madrid, Spain}
\affiliation{Departamento de Fisica de la Materia Condensada and Instituto "Nicol\'as Cabrera", Universidad Aut\'{o}noma de Madrid, 28049 Madrid, Spain}

\author{J. L. Fernandez Cu\~{n}ado}
\affiliation{IMDEA-Nanoscience, c/ Faraday, 9 Campus de Cantoblanco, Madrid, 28049 Madrid, Spain}
\affiliation{Departamento de Fisica de la Materia Condensada and Instituto "Nicol\'as Cabrera", Universidad Aut\'{o}noma de Madrid, 28049 Madrid, Spain}

\author{R. Guerrero}
\affiliation{IMDEA-Nanoscience, c/ Faraday, 9 Campus de Cantoblanco, Madrid, 28049 Madrid, Spain}

\author{M. A. Ni\~{n}o}
\affiliation{IMDEA-Nanoscience, c/ Faraday, 9 Campus de Cantoblanco, Madrid, 28049 Madrid, Spain}

\author{A. Bollero}
\affiliation{IMDEA-Nanoscience, c/ Faraday, 9 Campus de Cantoblanco, Madrid, 28049 Madrid, Spain}

\author{R. Miranda}
\affiliation{IMDEA-Nanoscience, c/ Faraday, 9 Campus de Cantoblanco, Madrid, 28049 Madrid, Spain}
\affiliation{Departamento de Fisica de la Materia Condensada and Instituto "Nicol\'as Cabrera", Universidad Aut\'{o}noma de Madrid, 28049 Madrid, Spain}
\affiliation{Condensed Matter Physics Center (IFIMAC), Universidad Aut\'{o}noma de Madrid, 28049 Madrid, Spain}

\author{J. Camarero}
\affiliation{IMDEA-Nanoscience, c/ Faraday, 9 Campus de Cantoblanco, Madrid, 28049 Madrid, Spain}
\affiliation{Departamento de Fisica de la Materia Condensada and Instituto "Nicol\'as Cabrera", Universidad Aut\'{o}noma de Madrid, 28049 Madrid, Spain}
\affiliation{Condensed Matter Physics Center (IFIMAC), Universidad Aut\'{o}noma de Madrid, 28049 Madrid, Spain}

\date{\today}
\begin{abstract}
We demonstrate that the interfacial exchange coupling in
ferromagnetic/antiferromagnetic (FM/AFM) systems induces
symmetry-breaking of the Spin-Orbit (SO) effects. This has been
done by studying the field and angle dependencies of anisotropic
magnetoresistance and vectorial-resolved magnetization hysteresis
loops, measured simultaneously and reproduced with numerical
simulations. We show how the induced unidirectional magnetic
anisotropy at the FM/AFM interface results in strong asymmetric
transport behaviors, which are chiral around the magnetization
hard-axis direction. Similar asymmetric features are anticipated
in other SO-driven phenomena.
\end{abstract}

 \maketitle

The Spin-Orbit (SO) interaction arises from the coupling of the
electron spin with its orbital motion~\cite{mcguire_IEEE_1975}.
SO effects influence both magnetic and transport properties and constitute 
the subject of modern nanomagnetism. The microscopic origin of the magnetic
anisotropy of ferromagnetic (FM) systems ultimately arises from
SO~\cite{fernando_book2008}, dictating the preferential
magnetization directions. In FM/heavy metal structures,
interfacial SO promotes a perpendicular magnetic anisotropy
(PMA)~\cite{storJMMM1999} and it is responsible of chiral spin
reversals~\cite{bogdanov_PRL_2001}, due to the
Dzyaloshinskii-Moriya interaction (DMI)~\cite{Fert_DMI}.
The SO interaction is exploited nowadays in spintronic
applications~\cite{fert_Nat2007}, since it produces a mixing of
the electron spin-up and spin-down states determining anisotropic
magnetoresistive signals. In addition, SO-induced spin Hall
effects may be exploited to efficiently manipulate and sense the 
magnetization in future spin-orbitronic applications~\cite{axel_IEEE2013}. 
In any case, the transport phenomena are strongly influenced by the 
effective symmetry of the SO effects. Therefore, determining their general
features represents a crucial step towards the understanding and the 
improvement of their functionalities.

In FM films with (two-fold) uniaxial magnetic anisotropy ($K_{\rm
U}$) SO determines symmetric magnetoresistance (MR) responses
around the magnetization easy-axis (e.a.) and hard-axis (h.a.)
directions~\cite{pernaAPL_2014}:
MR$(\alpha,\bf{H})=$MR$(-\alpha,\bf{H})=$MR$(\alpha,-\bf{H})$,
where $\alpha$ is the angle of the external applied magnetic field
$\bf{H}$ with respect to the anisotropy direction. This is the
anisotropic magnetoresistance (AMR), which depends on the angle
$\theta$ enclosed by the magnetization vector ($\textbf{M}$) and
the injected electrical current ($\textbf{J}$) following a
$\cos^2\theta$ law. Uniaxial systems present symmetric
magnetization reversal pathways and hence symmetric MR responses.
In this sense, a magnetic symmetry-breaking could promote
non-symmetric reversals and MR responses. For instance, a FM layer
exchange coupled with an antiferromagnet (AFM) layer presents an
additional (one-fold) unidirectional magnetic anisotropy ($K_{\rm
EB}$)~\cite{nogues_JMMM1999}, which is generally revealed through a
shift of the hysteresis loop of the FM layer, called exchange-bias
(EB) field, and an enhancement of the coercivity. From a
technological point of view, EB is largely exploited in
spintronics because it satisfies the need for stable and
controlled MR outputs in magnetic recording, processing, and
sensing devices. Systematic studies have shown how this
interfacial exchange coupling modifies the magnetization reversal
pathways~\cite{camarero_prl2005,camarero_prb2009}. The transport
properties have also been studied, but only for several fixed
magnetic-field values and/or field
directions~
\cite{miller_APL1996,brems_prl2007,Gruyters_JAP2004,Mattheis_IEEE2002}.

\begin{figure}[bp]
        \begin{center}
        \includegraphics*[width= 85 mm]{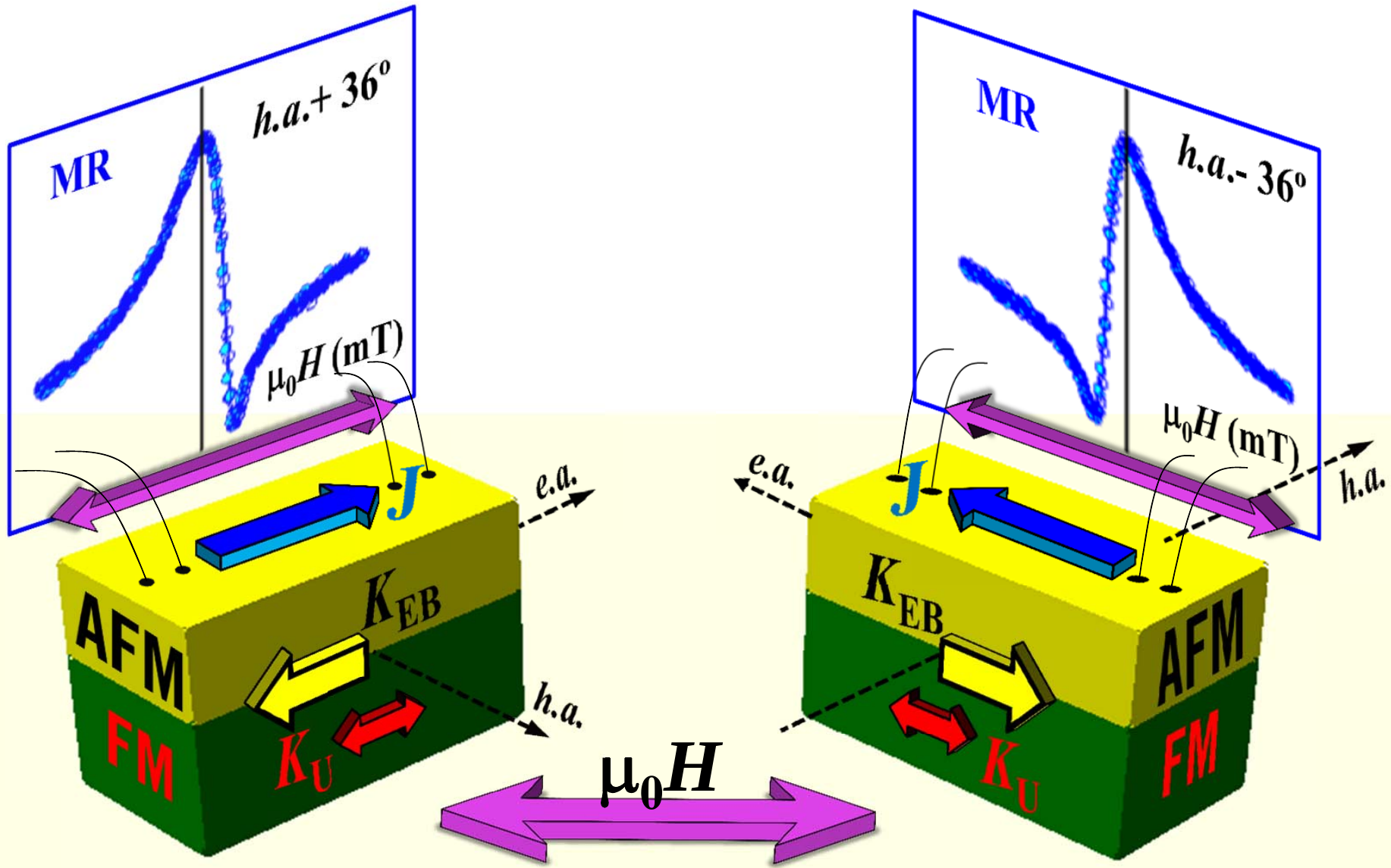}
            \caption [fig:figure1]{\label{figure1}
            Schematic representation of the chiral asymmetric transport
            behavior in FM/AFM system (as described in the text), which 
            arises from interfacial exchange coupling. The intrinsic (two-
            fold) $K_{\rm{U}}$ and
            interfacial-induced (one-fold) $K_{\rm{EB}}$ anisotropies are
            indicated with arrows. The top graphs display the corresponding
            MR curves acquired around the h.a.~direction.
            }
        \end{center}
\end{figure}

In this Letter we show that the interfacial exchange coupling in
FM/AFM systems strongly influences the SO effects thus promoting
asymmetric MR responses, which are chiral with respect to the
h.a.~direction, as Fig.~\ref{figure1} illustrates. Angular- and
field-dependent measurements of the MR and the magnetization
reversal pathways (measured simultaneously) have been reproduced
with numerical simulations without any free parameter. We show how
the symmetry of the MR response of an uniaxial FM system is broken
by the unidirectional exchange coupling imposed by an adjacent AFM
layer. In particular, symmetric MR responses are only found at
characteristic magnetization directions (e.a.~and h.a.).
Identical and chiral MR responses
are observed around the e.a.~and h.a.~direction, respectively.
Asymmetric magneto-transport behaviors produced by unidirectional symmetry-
breaking found in other spin-orbitronic systems are also discussed.

\begin{figure}[tp]
   \begin{center}
   \includegraphics*[width= 85 mm]{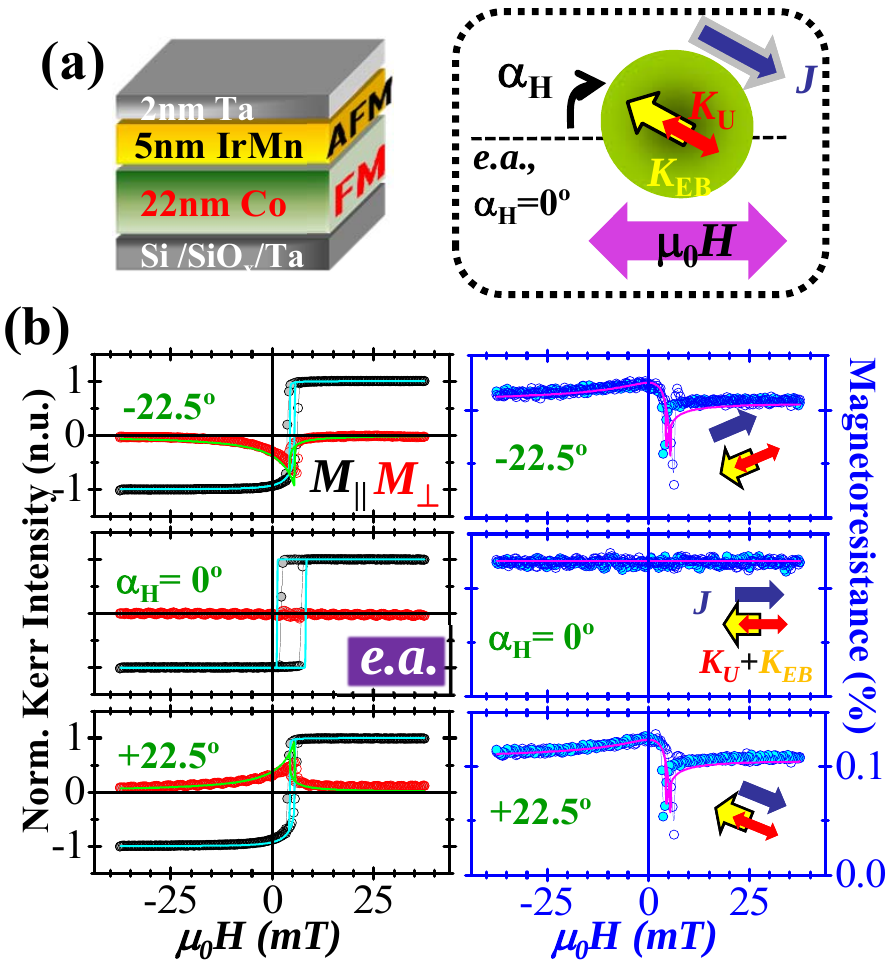}
      \caption [fig:figure2]{\label{figure2}
      (a) Schemes of sample structure (left) and experimental configuration 
      (right), indicating the directions of the anisotropies (collinear 
      $K_{\rm{U}}$ and $K_{\rm{EB}}$), current ($\bf{J} \parallel K_{\rm{U}}
      $), sample angle ($\alpha_{\rm{H}}$), and external magnetic field ($
      \mu_{0}H$).
      (b) Magnetic and transport study at selected $\alpha_{\rm{H}}$
      around the e.a. direction. The left column panel shows the $M_{\|}/
      M_{\rm{S}}$ and $M_{\bot}/M_{\rm{S}}$ hysteresis field loops whereas 
      the right column panel displays the corresponding MR loops. Symbols 
      are the experimental data and the continuous lines are the 
      corresponding simulated curves derived from the model described in the 
      text. Filled (empty) symbols refers to descending (ascending) branch. 
      The insets show schematically the specific current-anisotropy-field configuration. Notice
      the identical behavior of the MR(${\rm e.a.}\pm 22.5^{\circ})- H$ 
      curves.
      }
  \end{center}
\end{figure}

The scheme of the FM/AFM sample structure and
the experimental configuration are shown in Fig.~\ref{figure2}(a).
Details on the fabrication of the FM/AFM bilayers and reference FM films, 
with collinear uniaxial $K_{\rm{U}}$ and unidirectional $K_{\rm{EB}}$ are 
reported in Suppl.~Material~\cite{supplemental}.
Here, we refer to 22 nm Co/5 nm IrMn bilayer because it presents the smallest anisotropy field~\cite{supplemental}.

The magnetic and transport properties were studied at RT by
investigating the angular- and field-dependence of magnetization
reversal pathways and magnetoresistive responses.
Vectorial-resolved magnetization and resistance signals were
acquired \emph{simultaneously} as a function of the magnetic field
for a given field orientation
($\alpha_{\rm{H}}$)~\cite{pernaAPL_2014}. $\alpha_{\rm{H}}=0^{\circ}$
refers when the anisotropy axis is oriented parallel to the
external field (Fig.~\ref{figure2}(a)). The magnetization
components, parallel ($M_{\|}$) and perpendicular ($M_{\bot}$) to
the external field, were derived from vectorial-resolved magneto
optic Kerr effect measurements~\cite{supplemental,MOKE,perna_jap11}. The
magnetoresistance (MR) was measured by using a lock-in amplifier
in a four probe method with the electrical current vector set
parallel to the anisotropy axis (see Sec.~II in \cite{supplemental}). The measurements were performed in the whole angular range. In general, the
magnetization reverses via sharp irreversible (and smooth
reversible) transitions, indicative of nucleation and propagation
of magnetic domains (magnetization rotation)~\cite{perna_NJP}. The
relevance of this refers to the proximity to the e.a.~(h.a.) direction.
Consequently, MR depends strongly on $\alpha_{\rm{H}}$.

\begin{figure}[tp]
   \begin{center}
   \includegraphics*[width= 85 mm]{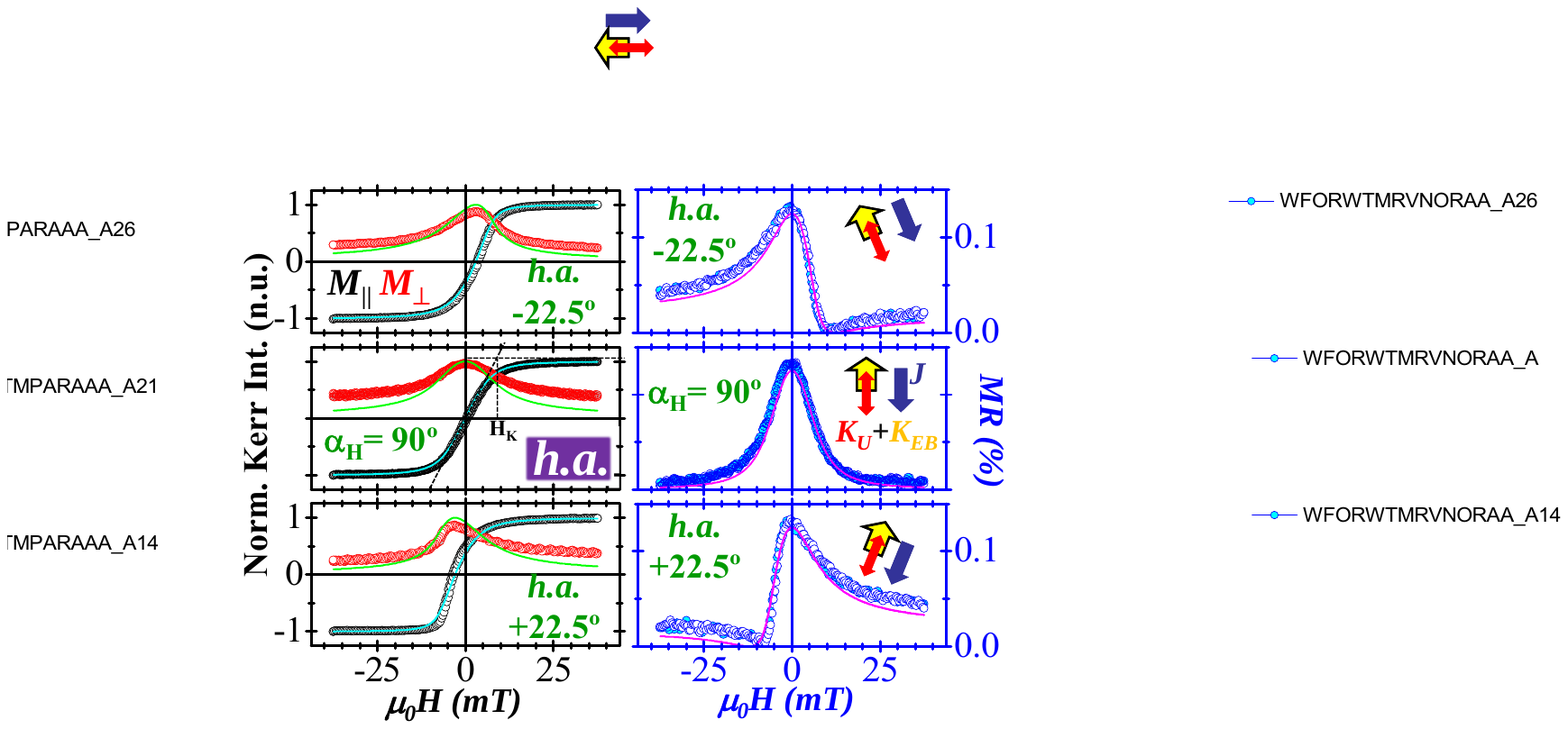}
      \caption [fig:figure3]{\label{figure3}
      Magnetic and transport study at selected $\alpha_{\rm{H}}$
      around the h.a.~with relevant magnetization rotation reversal. The
      left column panel shows the $M_{\|}/M_{\rm{S}}$
      and $M_{\bot}/M_{\rm{S}}$ hysteresis field loops whereas the right 
      column panel displays the corresponding MR loops. Symbols are the 
      experimental data and the continuous lines are the corresponding 
      simulated curves derived from the model described in the text. Filled 
      (empty) symbols refers to descending (ascending) branch. The insets 
      show schematically the specific current-anisotropy-field 
      configuration. Note the chiral asymmetry behavior of the MR(${\rm 
      h.a.}\pm 22.5^{\circ})- H$ curves.
      }
  \end{center}
\end{figure}

\begin{figure}[bp]
\resizebox{8.5cm}{!} {\includegraphics[scale=1.0]{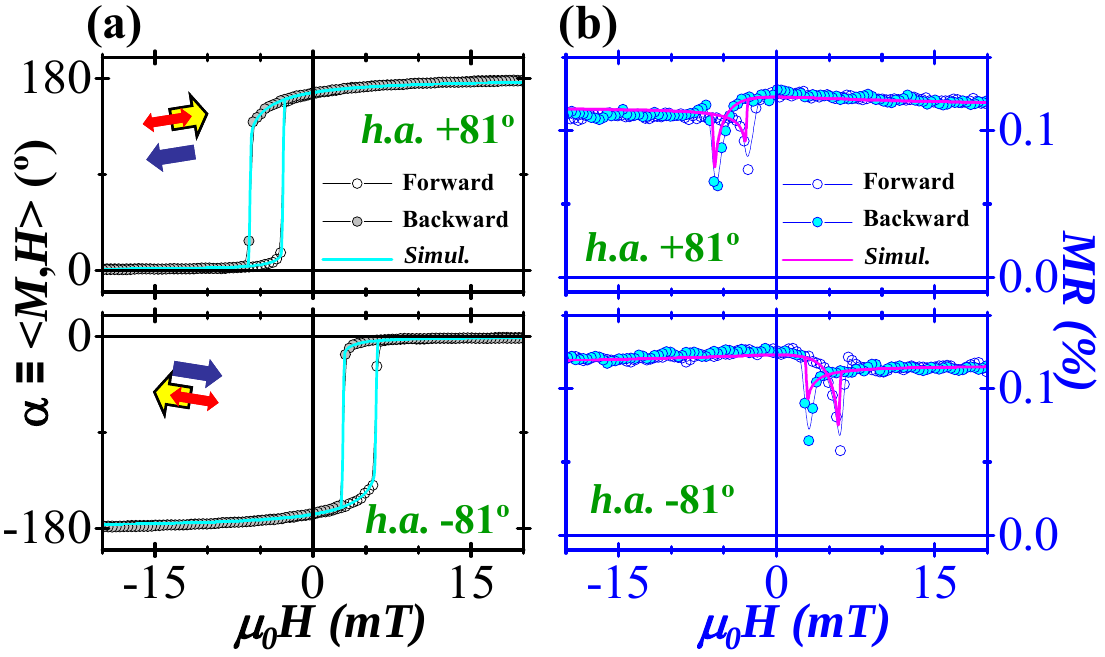}}
    \caption{(Color online)
      Magnetic and transport study at selected $\alpha_{\rm{H}}$ around the 
      h.a.~with relevant domain nucleation and propagation reversal. The
      left column panel shows the hysteresis field loops of the angle of the 
      magnetization vector ($\alpha\equiv\widehat{(\textbf{M},\textbf{H})}$) 
      whereas the right column panel displays the corresponding MR loops. 
      Symbols are the experimental data and
      the continuous lines are the corresponding simulated curves derived 
      from the model described in the text. Filled (empty) symbols refers
      to descending (ascending) branch. The insets show schematically
      the specific current-anisotropy-field configuration. Note
      the chiral asymmetry behavior of the MR(${\rm h.a.}\pm 81^{\circ})- H$ 
      curves.}
    \label{figure4}
\end{figure}

\begin{figure}[tp]
\resizebox{8.5cm}{!}{\includegraphics[scale=1.0]{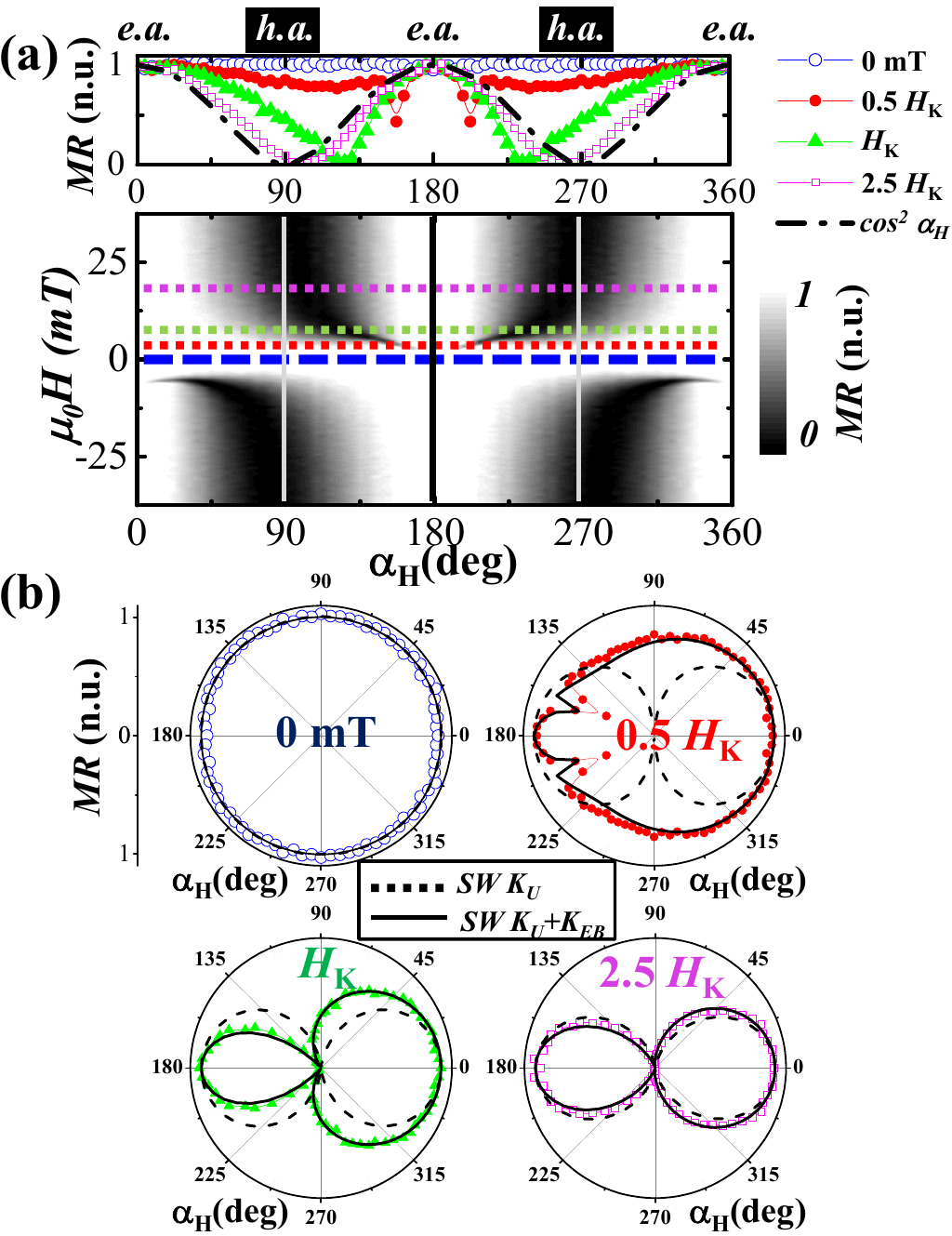}}
    \caption{(Color online) (a) Bottom graph: 2D field-angular map
    representation of the magnetoresistance MR derived from the
    forward field-branch hysteresis curves acquired at different
    angles, as the ones shown in the right panels of Fig.~\ref{figure2}, 
    Fig.~\ref{figure3} and Fig.~\ref{figure4} (similar information can be 
    derived from the backward 2D map~\cite{supplemental}). 
    The h.a.~(e.a.) directions are indicated with vertical
    grey (black) continuous lines. The dashed lines indicates the
    corresponding horizontal cuts at selected fields displayed in the top 
    graph. Notice that the MR angular dependence is approaching a 
    $cos^2\alpha_{\rm H}$ law (represented with a dotted-dashed curve) for 
    fields much larger than the anisotropy field $H_{\rm{K}}$.
    (b) Polar-plot representation of the angular dependence of the MR for
    different fields. Symbols are the experimental data. Solid
    (dashed) curves are derived from numerical simulation for the
    FM/AFM bilayer (single FM reference film). Note that both come closer at 
    high fields whereas at low fields they are very different. Remarkably, 
    for the (unidirectional) exchange-biased system reproduce the chiral 
    asymmetry around the h.a. direction. }
    \label{figure5}
\end{figure}

The correlation between magnetic and transport properties and the
general trends are determined from the comparison of the symmetry
relationships between them. This will be discussed in detail in
the following, first by comparing field-dependent curves at
selected angles around the characteristic e.a.~(Fig.~\ref{figure2}(b)) and 
h.a.~(Fig.~\ref{figure3} and
Fig.~\ref{figure4}) directions and then by comparing
angle-dependent curves at selected fields (Fig.~\ref{figure5}). It
is worth to remark that the high symmetry found in the
magnetotransport properties of a FM layer with uniaxial magnetic
anisotropy is no longer satisfied in the FM/AFM system. At first
glance, symmetric features in both magnetic and transport
properties are found only at the easy and hard direction. In any other
angular conditions, both magnetic and MR curves are strongly
asymmetric.

Fig.~\ref{figure2}(b) compares representative vectorial-resolved
magnetization (left panels) and magnetoresistance (right panels)
hysteresis loops acquired simultaneously at selected
$\alpha_{\rm{H}}$ around the e.a. At
$\alpha_{\rm{H}}=0^{\circ}$ (central left panel), $M_{\|}$-$H$
presents a shifted ($\mu_{\rm{0}}H_{\rm{EB}}=+4.7$ mT) squared shape
hysteresis loop with a sharp irreversible jump at $7.2$~mT, whereas the 
$M_{\bot}$-$H$ is
negligible in the whole field loop. Away from the e.a.,
$M_{\bot}(H)\neq 0$ and reverses only in one semicircle, and above
a critical angle, which depends on the ratio
$K_{\rm{U}}/K_{\rm{EB}}$~\cite{camarero_prl2005}, the
magnetization reversal becomes fully reversible. Therefore, close
to e.a.~direction nucleation and propagation of magnetic domains
are the relevant processes. The right panels in
Fig.~\ref{figure2}(b) display the corresponding transport
measurements. At the e.a.~the MR-$H$ curve is symmetric (and flat)
in the whole field loop, whereas non-symmetric curves are found
for $\alpha_{\rm{H}}\neq0^\circ$. As will be discussed below,
similar features are found for $M_{\|}$-$H$ and MR-$H$ around the
e.a.~direction.

Fig.~\ref{figure3} shows a similar study close to the h.a.
In general, both $M_{\|}$ and $M_{\bot}$ loops show
smooth reversible transitions, indicating that magnetization
rotation is the relevant process during reversal. $\textbf{M}$
rotates in-plane only in one semicircle during the reversal, so
that the angle between $\textbf{M}$ and $\textbf{J}$ is
continuously changing as the field is sweeping. At the h.a.~
(central panels of Fig.~\ref{figure3}), $M_{\|}$ displays a nearby
linear behavior, with an anisotropy field
$\mu_{\rm{0}}H_{\rm{K}}=9$~mT (see Sec.~III of \cite{supplemental}), whereas 
MR shows the maximum variation, which yields to $0.13\%$. In addition, the 
$M_{\|}$-$H$ curve show rotational symmetry whereas $M_{\bot}$-$H$ and MR-$H
$ curves are mirror symmetric with respect to zero field. However,
these symmetric features are lost away from the h.a.~direction.

Around the characteristic directions different
symmetry relationships are identified. For instance, for
$\alpha_{\rm{H}}=\pm22.5^{\circ}$, that is around the e.a.~direction
(Fig.~\ref{figure2}(b)), $M_{\|}$ and MR loops display identical
field dependent evolution, i.e.,
$M_{\|}(\alpha_{\rm{H}},H)=M_{\|}(-\alpha_{\rm{H}},H)$ and
MR$(\alpha_{\rm{H}},H)=$MR$(-\alpha_{\rm{H}},H)$, whereas the
$M_{\bot}$ experiences a sign change
$M_{\bot}(\alpha_{\rm{H}},H)=-M_{\bot}(-\alpha_{\rm{H}},H)$. In
turn, around the h.a.~direction, e.g., for
~$\alpha_{\rm{H}}=90^{\circ}\pm22.5^{\circ}$ (Fig.~\ref{figure3}),
the hysteresis curves of the parallel component are identical
under rotation around the origin, i.e.,
$M_{\|}(\rm{h.a.}+22.5^{\circ},H)=-M_{\|}(\rm{h.a.}-22.5^{\circ},-H)$,
whereas the hysteresis curves of the perpendicular component and
the MR display chiral asymmetry, i.e.
$M_{\bot}(\rm{h.a.}+22.5^{\circ},H)=M_{\bot}(\rm{h.a.}-22.5^{\circ},-H)$
and ${\rm MR}(\rm{h.a.}+22.5^{\circ},H)={\rm
MR}(\rm{h.a.}-22.5^{\circ},-H)$. This (2-dimensional) chiral asymmetry is
schematically illustrated in (Fig.~\ref{figure1}).

The chiral asymmetry is not only found closely around h.a., where
the magnetization reversal is governed by magnetization rotation
processes, but is extended in the whole angular range.
Fig.~\ref{figure4} shows the magnetic and transport behaviors for
two selected field directions around the h.a.,
$\alpha_{\rm{H}}=\rm{h.a.}\pm81^{\circ}$, but very close to the
e.a.~direction, where reversal is governed by nucleation and
propagation of domains. In this case, Fig.~\ref{figure4}(a) displays
the hysteresis curves of the angle $\alpha$ of the magnetization vector
with respect to the current direction extracted directly from the
vectorial-Kerr data. This angle defines the magnetic torque. The
corresponding resistance changes are shown in
Fig.~\ref{figure4}(b). In general, the MR loops present
pronounced MR peaks meanwhile the magnetization switches from
$\approx 0^{\circ}$ to $\approx \pm180^{\circ}$. There is an
asymmetry between the forward (descending) and backward
(ascending) field branches in both magnetic and transport
behaviors. This originates from the interfacial-induced
unidirectional anisotropy which results with more rounded
transitions and with higher MR peaks when the field sweeps against
the unidirectional
anisotropy~\cite{camarero_prl2005,brems_prl2007}. Moreover, the
chiral asymmetry is preserved, i.e.,
MR(h.a.$+81^{\circ},H$)=MR(h.a.$-81^{\circ},-H$. This indicates
that this asymmetry is independent of the reversal mechanism.

To gain further insight into the symmetry-breaking of the SO
effects, we have performed numerical simulations by using a
modified coherent rotation Stoner-Wohlfarth (SW) model (with no
free parameters) in which we included collinear uniaxial and
unidirectional anisotropy terms with
$K_{\rm{U}}/K_{\rm{EB}}=0.37$ (see Secs.~III and IV of \cite{supplemental}). 
This allows us to simulate the angular and magnetic field
dependence of the magnetization reversal
pathways~\cite{camarero_prb2009}, i.e., $M_{\|}(\alpha_H,H)$,
$M_{\bot}(\alpha_{\rm{H}},H)$, and to derive the corresponding MR
responses according to MR$(\alpha_{\rm{H}},H)\propto \cos^2
\theta(\alpha_{\rm{H}},H)$, where
$\theta\equiv\widehat{(\textbf{M},\textbf{J})}$. The simulated
hysteresis curves are superimposed (continuous black lines in
Figs.~\ref{figure2}(b), \ref{figure3} and \ref{figure4}) to the
experimental curves. There is a perfect agreement between them,
including their asymmetries, which demonstrates that both
magnetization pathways and magnetoresistance responses are
strongly affected by the system symmetry.

The broken-symmetry of the SO effects can be clearly observed by
plotting the whole angular evolution of the MR-$H$ hysteresis
loops in a 2D map representation (Fig.~\ref{figure5}(a)). Such
plot allows visualizing the broken periodicity. While the
well-defined uniaxial (two-fold) magnetic anisotropy of a FM
results with a $180^\circ$ periodicity~\cite{pernaAPL_2014}, the
additional unidirectional (one-fold) anisotropy induced at the
FM/AFM interface promotes the symmetry-breaking of the SO
effects~\cite{supplemental}, resulting with a $360^\circ$
periodicity.

In order to visualize the angle dependence of MR, different
horizontal cuts of the 2D map have been plotted in the top graph
in Fig.~\ref{figure5}(a). This represents the angular evolution
of MR at different magnetic field values. At remanence (i.e.,
$\mu_{0}H=0$ mT, blue empty-circles) the MR signal comes out from
a $\textbf{M}\|\textbf{J}$ configuration in the whole angular
range and therefore it does not change. For non-zero external
field, the angular dependence of the MR is clearly asymmetric, as
the mirror symmetry along the horizontal axis is completely
broken. From a simple inspection of the top graph of
Fig.~\ref{figure5}(a), we can figure out relevant information.
First, the maximum values of the MR are found when $\alpha_{\rm{H}}$ is
aligned along the anisotropy direction. $\textbf{M}$ is parallel
to $\textbf{J}$ only in a small range of angles, which increases
as the external field increases. Away from the anisotropy direction,
the MR value decreases gradually and it is minimum where the angle
between $\textbf{M}$ and $\textbf{J}$ approaches to $90^{\circ}$.
Second, larger MR changes are found as the external field
increases. Third, the MR displays identical or different (asymmetry) values 
around the anisotropy or h.a.~direction, respectively. The asymmetry 
vanishes for very large external magnetic fields compared to the anisotropy 
field, where the angle dependence of MR approaches a $\cos^{2}\alpha_{\rm{H}}$ law.

The discussed trends and symmetry-breaking SO effects are also
reflected in the corresponding polar plot representation of the MR
shown in Fig.~\ref{figure5}(b). Each graph includes both
experimental (symbols) and simulated (lines) data. The latter
includes model uniaxial (SW $K_{\rm u}$) and exchange-biased (SW
$K_{\rm u}+K_{\rm EB}$) systems. In remanence conditions, both
models are indistinguishable, i.e., circular-shaped polar plot with
no MR variation. For $\mu_0 H\neq 0$, the simulated polar plots of
the uniaxial system display a similar two-lobe behavior, which
result with mirror symmetry with respect to both e.a.~and h.a.~directions.
In contrast, the simulated polar plots of the
exchange-biased system show mirror symmetry with respect the e.a.~direction
and a (chiral) asymmetry with respect to the h.a.~direction.
The asymmetry diminishes as the external field
increases, vanishing for very large fields compared with the
anisotropy field. Remarkably, all experimental data are very well
reproduced by the model (see also Sec.~IV of \cite{supplemental}).

Asymmetric transport and magnetic behaviors, originating from interfacial 
symmetry-breaking (unidirectional) effects, can also be found in other 
magnetic systems.
For instance, exchange-biased spin valves display chiral symmetry in both
magnetization and giant magnetoresistance loops around the
h.a.~direction (see graph panels of Fig.~2 in
\cite{pernaPRB_2012}), as well as asymmetric MR curves were found
in exchange-biased multiferroic BiFeO$_{3}$-based
systems~\cite{allibe_nanoLett2012}. In addition, fixed chiral spin
structures can be stabilized in PMA systems with sizeable
DMI~\cite{chen_PRL2013}, and in the presence of an in-plane magnetic
field, producing asymmetric magnetization reversal features during
(chiral) magnetic domain nucleation~\cite{pizzini_PRL2013} and
domain propagation~\cite{hrabec_PRB2014}. On the other hand, in
FM/HM systems with in-plane magnetization, SO-dependent effective
in-plane field has been demonstrated giving rise to asymmetric
reversal~\cite{fan_natcomm2013}. Asymmetric planar
Hall~\cite{margin_scierep2015} and inverse spin
Hall~\cite{hammel_PRB2014} effect signals have also been recently
reported in exchange-biased insulating FM/metallic AFM bilayers,
as well as asymmetries in the spin Hall magnetoresistance has been
found in SrMnO$_3$/Pt~\cite{pan_PRB2013}. These asymmetric
signals, which are generally clearly observed at low-magnetic
fields, arise from the SO coupling in presence of an
unidirectional magnetic anisotropy. In the large field regime the
asymmetries vanish, in agreement with our results.

In summary, we have demonstrated the existence of broken-symmetry of the SO effects due to the exchange interaction
at the FM/AFM interface, which is responsible of asymmetries in
both magnetic and transport properties. In particular, we have
shown an intrinsic chiral asymmetry in the MR with respect to the
magnetization hard-axis direction. Similar effects can be
envisaged for other both spintronic and spin-orbitronic systems
when either intrinsic magnetic anisotropy or the SO interaction
presents symmetry-breaking.

This work was supported in part by the Spanish MINECO through
Project No.~MAT2012-39308, FIS2013-40667-P, MAT2011-25598 and MAT2014-52477-C5-3-P, and by the 
Comunidad de Madrid through Project S2013/MIT-2850 (NANOFRONTMAG-CM).
P.P.~and A.B.~acknowledge support through the Marie Curie AMAROUT EU 
Programme, and through "Juan de la Cierva" (JCI-2011-09602) and the "Ram\'{o}n y Cajal" contract from the Spanish MICINN, respectively.



\begin{thebibliography}{28}

\bibitem{mcguire_IEEE_1975} T. R. McGuire \& R. I. Potter, 
\emph{IEEE Trans. Magn.} \textbf{11}, 1018 (1975).

\bibitem{fernando_book2008} G. Fernando, 
\emph{Metallic Multilayers and their Applications}, edited by
Prasanta Misra, included in series Handbook of Metal Physics,
\emph{Elsevier} ISBN: 978-0-444-51703-6 (2008).

\bibitem{storJMMM1999}
J. St{\"o}hr, J. Magn Magn. Mater. \textbf{200}, 470 (1999).

\bibitem{bogdanov_PRL_2001} A. N. Bogdanov and U. K. R{\"o}{\ss}ler, \emph{Phys. Rev. Lett.} \textbf{87}, 037203 (2001). 

\bibitem{Fert_DMI} A. Fert, \emph{Mater. Sci. Forum} \textbf{5960}, 439480 (1990); A. Fert, and P.M. Levy, \emph{Phys. Rev. Lett.} \textbf{44}, 1538 (1980); A. Fert, V. Cros, and J. Sampaio, \emph{Nature Nanotech.} \textbf{8}, 152-156 (2013). 

\bibitem{fert_Nat2007} C. Chappert, A. Fert \& F. Nguyen Van Dau, 
\emph{Nature Mater.} \textbf{6}, 813 (2007).

\bibitem{axel_IEEE2013} A. Hoffmann, \emph{IEEE Trans. Magn.} \textbf{49}, 5172 (2013). 

\bibitem{pernaAPL_2014}
P. Perna, D. Maccariello, C. Rodrigo, J.L.F. Cu\~{n}ado, M.
Mu\~{n}oz, J.L. Prieto, M.A. Ni\~{n}o, A. Bollero, J. Camarero,
and R. Miranda, \emph{Appl. Phys. Lett.} \textbf{104}, 202407
(2014).

\bibitem{nogues_JMMM1999} J. Nogues and I. K. Schuller, J. Magn. Magn. Mater. \textbf{192}, 203 (1999).

\bibitem{camarero_prl2005} J. Camarero, \emph{et al.}, 
\emph{Phys. Rev. Lett.} \textbf{95}, 057204 (2005).

\bibitem{camarero_prb2009} E. Jim\'{e}nez, \emph{et al.}, 
\emph{Phys. Rev. B} \textbf{80}, 014415 (2009).

\bibitem{miller_APL1996} B. H. Miller and E. Dan Dahlberg,  \emph{Appl. Phys. Lett.} \textbf{69}, 3932 (1996) 

\bibitem{brems_prl2007} S. Brems, K. Temst, and C. Van Haesendonck,
\emph{Phys. Rev. Lett.} \textbf{99}, 067201 (2007).

\bibitem{Gruyters_JAP2004} M. Gruyters, \emph{J. Appl. Phys.} \textbf{95}, 2587 (2004). 

\bibitem{Mattheis_IEEE2002} K.-U. Barholz and R. Mattheis, \emph{IEEE Trans. Magn.} \textbf{38}, 2767 (2002). 

\bibitem{supplemental}
Supplementary information available at http://xxxxxx provides detailed information on experimental/simulation procedures and extended data.

\bibitem{MOKE} E. Jim\'{e}nez, \emph{et al.}, 
Rev. Sci. Instrum. \textbf{85}, 053904 (2014); J. L. Cu\~{n}ado, \emph{et al.},
Rev. Sci. Instrum. \textbf{86}, 046109 (2015).

\bibitem{perna_jap11} P. Perna, \emph{et al.}, 
    \emph{J. Appl. Phys.} \textbf{110}, 13919 (2011); \emph{ibid.} 
    \emph{J. Appl. Phys.} \textbf{109}, 07B107 (2011).

\bibitem{perna_NJP} P. Perna, L. M\'{e}chin, M. Sa{\"i}b, J. Camarero and S. Flament, \emph{New J. Phys.} \textbf{12}, 103033 (2010).

\bibitem{pernaPRB_2012}
P. Perna, C. Rodrigo, M. Mu\~{n}oz, J. L. Prieto, A. Bollero, D.
Maccariello, J. L. F. Cu\~{n}ado, M. Romera, J. Akerman, E.
Jim\'{e}nez, N. Mikuszeit, V. Cros, J. Camarero, and R. Miranda,
\emph{Phys. Rev. B} \textbf{86}, 024421 (2012).

\bibitem{allibe_nanoLett2012} J. Allibe, S. Fusil, K. Bouzehouane, C. Daumont, D. Sando, E. Jacquet, C. Deranlot, M. Bibes, and  A. Barthelemy, \emph{Nano Lett.} \textbf{12}, 1141 (2012). 

\bibitem{chen_PRL2013} G. Chen, J. Zhu, A. Quesada, J. Li, A.T. NDiaye, Y. Huo, T.P. Ma, Y. Chen, H.Y. Kwon, C. Won, Z.Q. Qiu, A.K. Schmid, and Y.Z. Wu, \emph{Phys. Rev. Lett.} \textbf{110}, 177204 (2013). 

\bibitem{pizzini_PRL2013} S. Pizzini, J. Vogel, S. Rohart, L. D. Buda-Prejbeanu, E. Ju\'{e}, O. Boulle, I. M. Miron, C. K. Safeer, S. Auffret, G. Gaudin, and A. Thiaville, \emph{Phys. Rev. Lett.} \textbf{113}, 047203 (2014). 

\bibitem{hrabec_PRB2014} A. Hrabec, N. A. Porter, A. Wells, M. J. Benitez, G. Burnell, S. McVitie, D. McGrouther, T. A. Moore, and C. H. Marrows, \emph{Phys. Rev. B} \textbf{90}, 020402(R) (2014). 

\bibitem{fan_natcomm2013} X. Fan, J. Wu, Y. Chen, M. J. Jerry, H. Zhang, and J. Q. Xiao, \emph{Nature Comms.} \textbf{4}, 1799 (2013). 

\bibitem{margin_scierep2015} X. Zhou, L. Ma, Z. Shi, W. J. Fan, R. F. L. Evans, Jian-Guo Zheng, R. W. Chantrell, S. Mangin, H. W. Zhang and S. M. Zhou, \emph{Sci. Rep.} \textbf{5}, 9183 (2015). 

\bibitem{hammel_PRB2014} C. Du, H. Wang, F. Yang, and P. C. Hammel, \emph{Phys. Rev. B} \textbf{90}, 140407(R) (2014). 

\bibitem{pan_PRB2013} J. H. Han, C. Song, F. Li, Y. Y. Wang, G. Y. Wang, Q. H. Yang, and F. Pan, \emph{Phys. Rev. B} \textbf{90}, 144431 (2014). 

\end{thebibliography}
\end{document}